# Survey of Parallel Computing with MATLAB

Zaid Abdi Alkareem Alyasseri
ITRDC - University of Kufa
Kufa, P.O. Box (21), Najaf Governarate, Iraq
zaid.alyasseri@uokufa.edu.iq

*Abstract* — **Matlab is one of the most widely used mathematical computing environments in technical computing. It has an interactive environment which provides high performance computing (HPC) procedures and easy to use. Parallel computing with Matlab has been an interested area for scientists of parallel computing researches for a number of years. Where there are many attempts to parallel Matlab. In this paper, we present most of the past,present attempts of parallel Matlab such as** *MatlabMPI, bcMPI, pMatlab, Star-P* **and** *PCT*. **Finally, we expect the future attempts.**

1. INTRODUCTION

The most of software has been written for serial computation. That means it is run on a single computer which having a single Central Processing Unit (CPU) .Therefore,the problem will be divided into a number of series instructions. Where the execution of the instructions will be sequentially [1].

Parallel computing is one of the computing methods which execute many computation (processes) simultaneously. Where the principle of parallel computing is often can be divided the large problem into smaller pieces, then solved that concurrently ("in parallel") [2] .

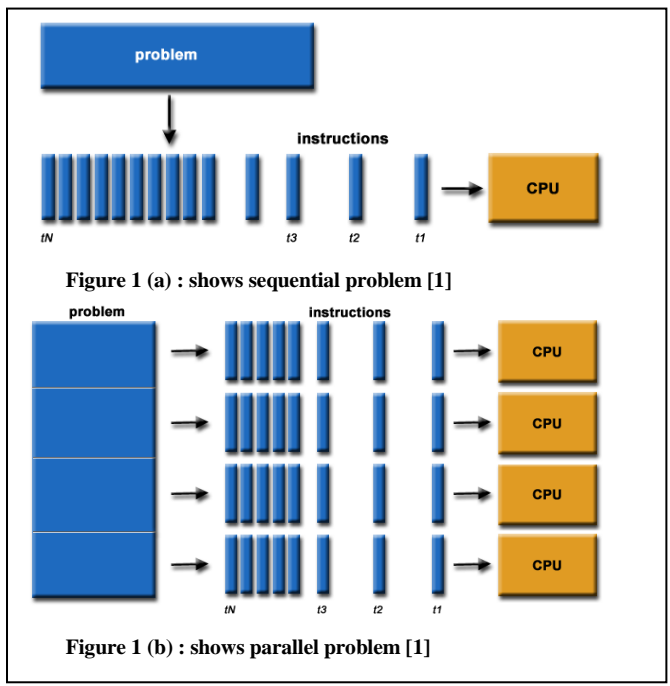

Figure 1 (a) : shows sequential problem [1]

Figure 1 (b) : shows parallel problem [1]

On other words, parallel computing is use of the multiple compute resources to solve a computational problem simultaneously . Which to be run on multiple CPUs. Figure 1 (a) and (b) shows how to divide the problem in sequential and parallel .

In fact, the main advantages of parallel computing are : 1) save time and/or money; 2) solve larger problems; 3) provide concurrency; 4) use of non-local resources; and 5) limits to serial computing [1],[2],[3].

*1.1 Computer Memory Architectures*

In general, main memory in parallel computer hardware can be classified into three kinds: *shared memory, distributed memory and distributed shared memory*. Shared memory is all processors interconnections with big logical memory (the memory is not physically distributed). Distributed memory refers to each processor has own local memory. Distributed shared memory combine the two previous approaches, where the processing element has its own local memory and access to the memory on non-local processors. Indeed, accesses to local memory are typically faster than accesses to non-local memory [1], [2], [3]. Figure 2 illustrates architectural differences between distributed and shared memory .

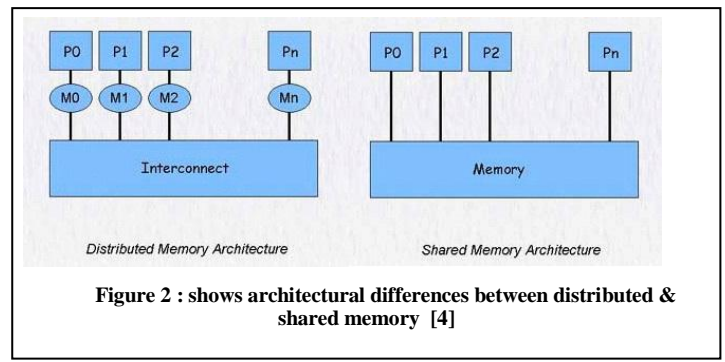

**Figure 2 : shows architectural differences between distributed & shared memory [4]**

Computer hardware architectures were developed rapidly for getting new benefits such as *power reducing, exploit all processors in modern computers that contain more than one processor* . On the other side, parallel computing software also evolved to achieve the advantages of parallel computing .One of the most important of these softwares is *Matlab* .Will address how Matlab language was developed to be compatible with parallel computing in the next section.

The paper is organized as follows: Section 2 presents the background on Matlab. Section 3 will be history of parallel Matlab and the various attempts to produce it. Section 4 summarizes the development of the parallel computing in Matlab. The current research in parallel computing with Matlab that will discuss in section 5. Section 6 will be parallel computing in Matlab for the future. Finally, Section 7 the conclusions and recommendations.

## 2. BACKGROUND ON MATLAB

*Matlab (matrix laboratory)*

*Matlab is* a numerical computing environment and fourth-generation programming language. Developed by **MathWorks** in 1984 and the newest version of Matlab is R2012b, Matlab allows matrix manipulations, plotting of functions and data, implementation of algorithms, creation of graphic user interfaces (GUI), and interfacing with programs written in other languages, including C, C++, Java, and Fortran[5],[6]. There are many advantages of programming with Matlab: 1) it's easy to learn and use where allowing non-experts to make changes and experiment; 2) it is fast and it has good supporting from (http://www.mathworks.com/) by tutorials and documents (*getting started*) and integrated help with built-in example code (*help library*); 3) it is interpretive and interactive and it has excellent debugging capabilities; and 4) Matlab is widely used in academia, industry (almost 1 million users by some estimates) [6],[7],[8]. But it is not free so you have to get licenses[7]. While Matlab is widely popular software used in many applications *such as image and signal processing, control systems, statistical estimations, among machine learning community, financial modeling, and computational biology*. Problems in these *fields* are computationally intensive requiring good processing power to solve the mathematical models within a reasonable period of time[8]. Where these applications need to be faster. Indeed, parallel programming using C/C++/FORTRAN and MPI is hard also creating parallel code in these languages takes a long time*[9]*. For these reasons, the scientists in **MathWorks** tried to apply the principle of parallel programming in MATLAB. In order to obtain benefits from the parallel computing to solve the big problems as *fast* and efficient. There are many attempts to develop parallel Matlab.

### 2.1 HISTORY OF PARALLEL COMPUTING WITH MATLAB HISTORICAL PERSPECTIVE (1995 TO 2011)

As previously mentioned, parallel Matlab has been an interested area for scientists of parallel computing researches for a number of years. There are many different approaches and attempts have been presented. Here will review the most important of these approaches.

The first attempt in this area for *Cleve Moler*, the original Matlab author and cofounder of The MathWorks, he has presented in **1995** "*why there isn't a parallel in MATLAB*"

[7], where the author described three challenges in developing a parallel Matlab language: *memory model, granularity of computations,* and *business situation.*

In that work, the author explained the most important attributes of a parallel computing like use shared or distributed memory. In this direction, Cleve Moler discussed the first attempt to make parallel Matlab in *iPSC*. It is one of the first commercially available parallel computers from Intel. The *iPSC* means "*Intel Personal SuperComputer*" 1985, it consisted of 32 to 128 nodes arranged in an ethernet-connected hypercube. Each node had a 80286 CPU with 80287 math coprocessor, 512K of RAM[7],[10]. Where each node could execute different program. One of the important points in Matlab which support memory model. There are no declarations or allocations because it is all handled automatically. The second challenge that was discussed in that work is *granularity*. The author presented an example of the *Ardent Titan*, where applied parallel Matlab on a shared memory multiprocessor. The third challenge is *business situation*, where Cleve decided to make some changes in Matlab's architecture, at the same time, the author won't expect the appearance of distinct improvements in parallel programming in Matlab in the near future.

After Cleve Moler, there are many attempts to parallelize Matlab, but all of them faced some problems. The most important problems that faced them are the parallel: 1) time in Matlab is not very attractive; 2) there are a few resources in Matlab; and 3) the biggest problem is there were not enough Matlab users who wanted to use Matlab on parallel computers, but they wanted more focused on improving the uniprocessor Matlab [7]. However, The MathWorks and several Institutions did not stop at this point, but go on to develop parallel Matlab.

With the passage of time, several factors have made the parallel Matlab as one of the most important project inside the software development institutes because Matlab has become more widely used and it presented easy access to multiprocessor machines. Consequently, there are three approaches to develop parallel Matlab: 1) *translating Matlab into a lower-level language; 2) using Matlab as a "browser" for parallel computations; and 3) extend Matlab through libraries or by modifying the language itself*[6],[8].

***The first approach*** is translated Matlab code into C / C++ or FORTRAN. Then will be parallelized the resulting code using one of parallel languages such as MPI or OpenMP, after that execute it on a HPC platform.

***The second approach*** is to use Matlab as a "browser" for parallel computations on a parallel computer. While Matlab still does not run as a parallel, in fact we cannot classify this method not more than a "web browser" to access to parallel applications. This approach used from Intel Hypercube and Ardent Titan[6]. Now a commercial project called Star-P (*P).

Both of these approaches have significant limitations such as: 1) it's expensive; 2) maybe it has an error-prone; 3) takes time; and 4) it is very difficult where change Matlab code into C/C++ or FORTRAN it not easy and at the same time the modify it will be more difficult because single line Matlab correspond many lines in C language [6],[7],[8].

As a solution to these limitations, *the third approach* came, extend Matlab through libraries or by modifying the language itself. In this paper will be more focused on this approach.

Consequently, third approach there are two methods to develop parallel Matlab: 1) writing *explicit* code to perform inter-processor communication at the HPC platform such as *MatlabMPI* and *bcMPI* ; 2) writing *implicit* code such as *pMatlab*, *StarP* and the Parallel or Distributed Computing Toolbox *(PCT)* or *(DCT)* [8]. We will describe all of them as following :

### 2.1.1 MatlabMPI

MatlabMPI is a pure Matlab script implementation based on basic MPI functions[8]. It designed by **Dr. Jeremy Kepner** in the Lincoln Laboratory at Massachusetts Institute of Technology(MIT) [11]. Where the MIT Lincoln Labs are an implementation it based on six MPI standard functions [12] . The functions required for MatlabMPI as follows:

| Function Name | Function Description |
|---|---|
| MPI_Init | Initializes MPI. |
| MPI_Comm_size | Gets the number of processors in a communication. |
| MPI_Comm_rank | Gets the rank of current processor within a communicator. |
| MPI_Send | Sends a message to a processor. |
| MPI_Recv | Receives a message from a processor. |
| MPI_Finalize | Finalizes MPI. |

Table 1: Selected MPI functions provided by MatlabMPI[12].

This method has many advantages such as its use just the Matlab language for implementation. Consequently, it can be run anywhere Matlab available. The second advantage of MatlabMPI is that applies MPI communication of standard Matlab file I/O. See to figure 3 that shows the communication system. Also, this approach has a very small library (~300 lines) and is highly portable. The price for this portability is that the while MatlabMPI performance is comparable to C+MPI for large messages, its latency for small messages is much higher (see Figure 4).

Finally, this approach has disadvantages,where requires a shared file system accessible by all processors. Where the system will face license problem for each node. So if its implements with shared memory machine only one Matlab license is required to run MatlabMPI. However, when MatlabMPI is run on a distributed system, such as a Linux cluster, a license for each multi CPU node is required[6]. As a solution for this limitation is to use intelligent compiler configuration tools for MatlabMPI that developed at Ohio Supercomputer Center in 2006 [13]. Where these tools presented enhance to MatlabMPI by convert the MatlabMPI scripts which do not need Matlab licenses into alone executable then run them in parallel[13].

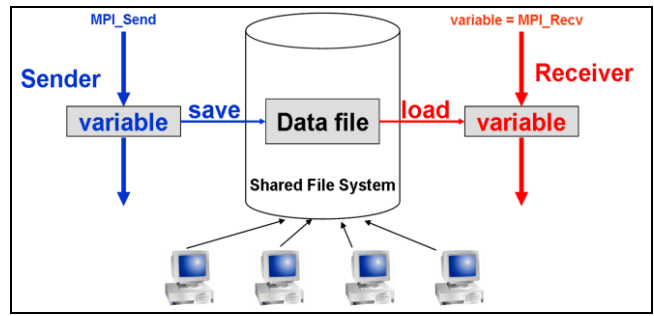

Figure 3: MatlabMPI file I/O based communication[14].

Figure 3 shows the communication(sending/receiving) of MatlabMPI. Where MatlabMPI applied the point-to-point communication non-blocking. The sender writes variables to a buffer file and then writes a lock file. The receiver waits until it sees the lock file, it then reads in the buffer file.

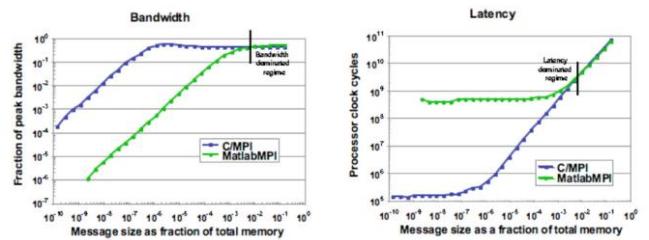

Figure 4: MatlabMPI *vs* C+MPI[14].

MatlabMPI and C+MPI in figure 4 were compared based on Bandwidth , latency and message size. Bandwidth is given as a fraction of the peak underlying link bandwidth. Latency is given in terms of processor cycles.For large messages the performance is comparable. For small messages the latency of MatlabMPI is much higher.

### 2.1.2 bcMPI

As we mentioned , there are several methods of parallel Matlab. The bcMPI is one of these methods. bcMPI is an open source software library, which is an alternative to MatlabMPI. bcMPI developed by the Blue Collar computing software at Ohio Supercomputer Center(OSC) in 2006 to provide a scalable communication mechanism and an efficient for development parallel Matlab also bcMPI is kept to compatible with MatlabMPI [9],[15]. According to,[8],[9],[15], there are several advantages in bcMPI such as : 1) it has core library and separate MPI toolboxes for each of these interpreters Matlab (*mexmpi*) and GNU Octave (*octmpi*) ; 2) bcMPI has ability to work with large problem like SIP problems; 3) it is compatible with MPI API ;4) bcMPI has supports a collection of Matlab and Octave data types; and 5) bcMPI has been developed primarily on the Linux platform, but it has also been tested on the Mac OS-X, NetBSD and IA32 platforms.

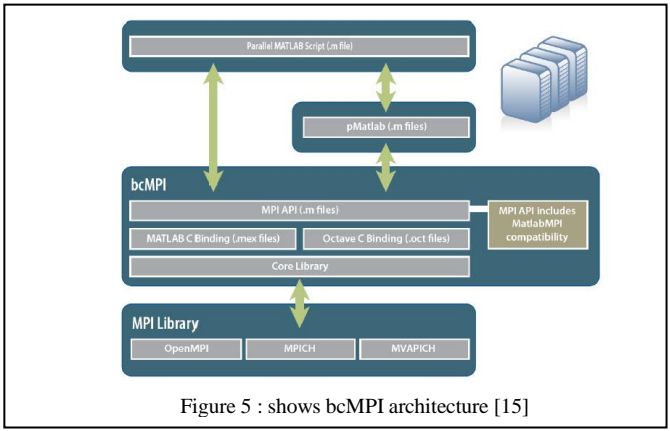

Figure 5 : shows bcMPI architecture [15]

Figure 5 illustrates the relationship between the various layers in the bcMPI architecture. Where the bcMPI has supported MPI functions and as summarize for MatlabMPI and bcMPI table1 shows the comparison between them.

TABLE I.     COMPARISON BETWEEN BCMPI AND MATLABMPI

| Product name / function | bcMPI | MatlabMPI |
|---|---|---|
| tag values | integer values , but reusable | Random numbers or strings, must be unique |
| collective tags | optional, ignored | required |
| MPI_Bcast() | compatible with MatlabMPI | root does MPI_Send() to each process, receivers may call MPI_Bcast() or MPI_Recv() |
| MPI_Broadcast() | efficient use of underlying MPI_Bcast() | not implemented |
| asynchronous send | may block, unless user specifies sufficient memory size with MPI_Buffer_attach() | never blocks, unlimited buffering using NFS files |
| MPI_Probe() | uses underlying MPI_Iprobe(), returns at most one pending message | returns all pending messages |
| MPI_Reduce() | supports any Matlab function with MPI_Op_create() | not implemented |

### 2.1.3 pMatlab

pMatlab developed by Lincoln Labs at Massachusetts Institute of Technology (MIT) in 2004[8] . It is defined as a parallel programming toolbox which based on PGAS (partitioned global address space) approach [16],[17]. The pMatlab implemented an implicit programming approach that gave several advantages such as: 1) It is supporting global arrays for optimized performance by enable integrating between global arrays and direct message passing ; 2) It does not need any external libraries, where all implemented occurs inside Matlab ; 3) that gave other good point is providing support for distributions and redistributions of up to four-dimensional arrays distributed with any combination of block-cyclic distributions; and 4) It is possible large data like SIP applications. Figure 5 shows the pMatlab layer architecture. pMatlab uses by default MatlabMPI communication library but for more efficient it can be used bcMPI[14].

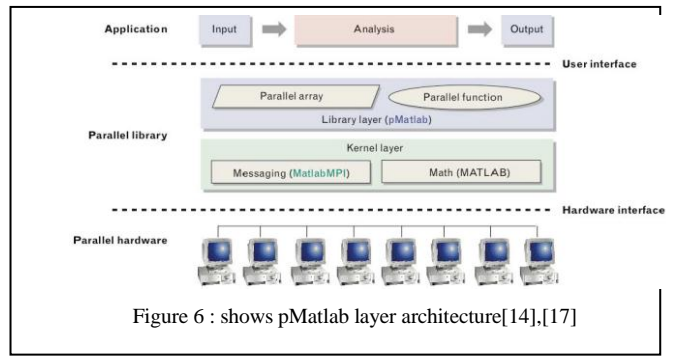

Figure 6 : shows pMatlab layer architecture[14],[17]

As we mentioned, one of the most important point in pMatlab is supporting global array. In fact pMatlab supports both pure and fragmented global array programming models. Where the pure global array programming model requires few changes in the code for providing the highest level of abstraction. In contrast, pMatlab in fragmented global array programming model provides guarantees on performance [14],[17].

*2.1.3.1 pMatlab benchmark*

Here will present results of four pMatlab benchmark ( ***STREAM, FFT, Top500* and *RandomAccess*** ) . The performances are compared with serial MATLAB and C+MPI. For implementation pMatlab and C+MPI was used cluster has 128 processors. In fact , the the C+MPI can run double size of the problem in pMatlab because pMatlab need to create temporary arrays when using high-level expressions .As result from figure 6 the pMatlab performance levels closer than C+MPI in FFT and STREAM , there is some slower in Top 500 but it is so slower in RandomAccess. However, the code size in pMatlab is smaller than C+MPI[14],[17].

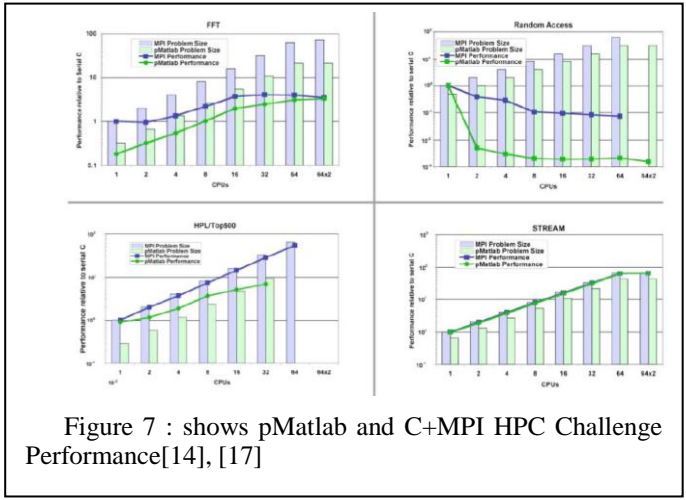

Figure 7 : shows pMatlab and C+MPI HPC Challenge Performance[14], [17]

### 2.1.4 Star-P

Another types of parallel Matlab is Star-P. Star-P Matlab developed at Interactive Supercomputing Corporation (ISC) in 2004 [8] . ISC launched in 2004 as part of MIT , in 2005 became an independent. Star-P Matlab is a set of extensions to Matlab. The main purpose of Star-P Matlab is to make the parallel of common computations are more simple. Where the Star-P categorize as client-server parallel computing platform in supercomputing , which has Matlab[18]. Figure 8 shows the architecture of Matlab Star-p where the Matlab will be as client and the High Performance Computing platform as server.

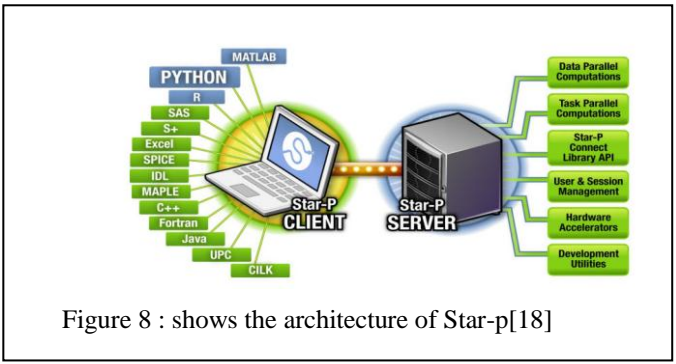

Figure 8 : shows the architecture of Star-p[18]

The approach of Star-P based on implicit programming where the kernel of star-P written using C++ language and MPI. The code of Star-P is like Matlab code just put **\*P** for telling Matlab this code use Star-P for instance :

**x=rand(2000\*p,2000\*p);**

When Matlab gets \*P, that means the next code uses Star-P type. The previous code will reserve **x** matrix 2-D (2000\*2000) elements as random values to the parallel processor memory space[8].

### 2.1.5 MATLAB DCT and *MATLAB PCT*

Parallel Computing Toolbox (PCT) is the software introduced from the Mathworks in November 2004,(originally named Distributed Computing Toolbox™ and Matlab Distributed Computing Engine™, respectively but after that divided into Parallel Computing Toolbox and Distributed Computing Toolbox ). Both Parallel and Distributed Computing Toolbox are based on the implicit programming approach[6],[8],[19] .

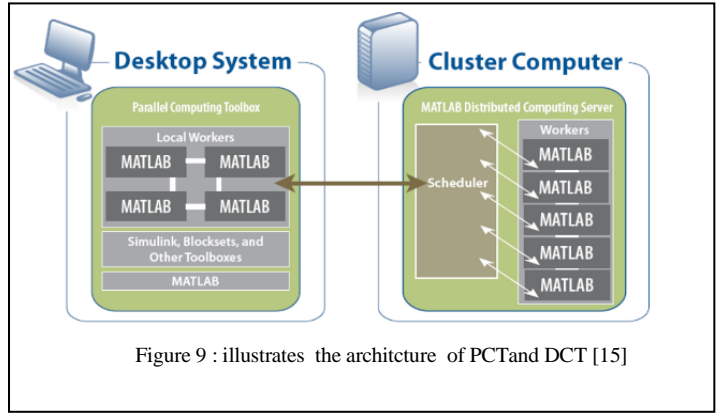

Figure 9 : illustrates the architcture of PCTand DCT [15]

Figure 9 – shows the architecture of PCT and DCT, where the tools started allowing users to run up to **4** Matlab Labs or Workers on single machine[19]. In the version R2009a allows users to run up 8 Matlab Labs on single machine [18]. The version *R2011b* allows to use 12 users on single machine[19].

In 2008 MathWorks launched issuing Parallel Computing Toolbox as a separate product on Distributed Computing Toolbox[6],[19].

The advantages of PCT are supporting high level constructs for instance parallel *for-loops* and *distributed arrays*, it's has a lot of *math functions*, and it provides users utilize existing Matlab[6] .

*Development of PCT*
According to, MathWorks [19],Table 2 shows stages of development in the Parallel Computing Toolbox (PCT) from 2006 until the first half of 2012 . For more details see[19].

| Matlab version | The new features of Parallel Computing Toolbox |
|---|---|
| **R2006b** | <ul><li>support Windows 64 (*Win64*) for both Matlab client and worker machines.</li><li>*Parallel Job Enhancements* where now it is become support any Scheduler for parallel jobs such as Windows Compute Cluster Server (CCS), Platform LSF, mpiexec,</li><li>*Improve error detection* for miscommunication between running parallel workers.</li><li>*Distributed arrays are partitioned into segments*, that means will be more efficient use of memory and faster, where each segment will send to a</li></ul> |

| | | |
|---|---|---|
| | | different lab. Finally, each lab will be processed and store the own part. |
| R2007a | | • Allowing users to run up to 4 Matlab Labs on single machine.<br>• New pmode interface and new default scheduler for pmode<br>• *Enhanced Matlab Functions*:here some Matlab functions have been enhanced to deal on distributed arrays such as : *cat,find,horzcat,subsindex and vertcat.*<br>• darray Function Replaces distributor Function<br>• rand Seeding Unique for Each Task or Lab<br>• Single-Threaded Computations on Workers |
| R2007b | | • New Parallel for-Loops (*parfor-Loops*)<br>• A new graphical user interface for creating and modifying user configurations.<br>• Parallel Profiler. |
| R2008a | | • Renamed functions for product name changes<br>• Changed function names for distributed arrays<br>• **Parallel Computing Toolbox support**<br>• *parfor* Syntax has single usage<br>• dfeval Now Destroys Its Job When Finished |
| R2008b | | • **Matlab compiler product support for Parallel Computing Toolbox applications.**<br>• Rerunning failed tasks.<br>• Enhanced job control with generic scheduler interface<br>• Composite objects provide direct access from the client (desktop) program to data that is stored on labs in the MATLAB pool. |
| R2009a | | • **A number of local workers increased to 8.**<br>• Admin center allows controlling of cluster resources like start, stop and otherwise control job.<br>• Support Microsoft Windows HPC Server 2008 (CCS v2).<br>• Pre-R2008b distributed array syntax. |
| R2009b | | • Renamed codistributor functions<br>• Enhancements to admin center<br>• Updated globalIndices Function<br>• Support for Job Templates and Description Files with HPC Server 2008.<br>• pctconfig Enhanced to Support Range of Ports<br>• Random Number Generator on Client Versus Workers |
| R2010a | | • Enhanced Functions for Distributed Arrays<br>• taskFinish File for Matlab Pool |
| R2010b | | • **GPU Computing.**<br>• Job Manager Security and Secure Communications<br>• Generic Scheduler Interface Enhancements<br>• Enhanced Functions for Distributed Arrays<br>• Support for Microsoft Windows HPC Server 2008 R2. |
| R2011a | | • **Enhanced GPU Support**<br>• Distributed Array Support<br>• Enhanced parfor Support.<br>• support Microsoft Windows HPC Server on 32-bit Windows clients.<br>• Enhanced Admin Center Support. |
| R2011b | | • Number of Local Workers **Increased to 12**.<br>• New Job Monitor<br>• **Enhanced GPU Support where R2011b only support the latest NVIDIA CUDA device driver.** |
| R2012a | | • Enhanced Distributed Array Support<br>• Task Error Properties Updated<br>• **New Programming Interface**<br>• Cluster Profiles<br>• **Enhanced GPU Support**<br>• Random Number Generation on Workers |

*Summary of history parallel Matlab*

Finally, as summarized for these methods table 2 shows the analysis of these methods.

TABLE II. ANALYSIS OF THE PREVIOUS ATTEMPTS IN MATALB PARALLEL

| Product name | Properties | | | |
|---|---|---|---|---|
| | Developed by | Year | Approach | Category |
| **Matlab MPI** | *MTI* | *2001* | *Explicit* | *message passing* |
| **pMatlab** | *MIT* | *2004* | *Implicit* | *partitioned global address space (PGAS)* |
| **Star-P** | *ISC* | *2004* | *Implicit* | *client/server* |
| **DCT** | *MathWorks* | *2004* | *Implicit* | *message passing* |
| **bcMPI** | *OSC* | *2006* | *Explicit* | *message passing,* |
| **PCT** | *MathWorks* | *2008* | *Implicit* | *message passing,* |

### 3. CURRENT RESEARCH IN PARALLEL MATLAB

In this section will be focused on Parallel Computing Toolbox in R2012b especially Matlab PCT with GPU. At the end of this section there is discussion some of benchmarks which solved using GPU and compare it with CPU.

*3.1 Graphics Processing Unit (GPU)*

Graphical Processing Units (GPUs) were invented in 1999 by NVIDIA. A GPU is a highly parallel computing device. It's designed to accelerate the analysis of the large datasets such as image , video and voice processing or to increase the performance with graphics rendering , computer games[21]. In the last ten years, the GPU has a major development where it became used in many applications such as the iterative solution of PDEs, video processing, machine learning, and 3D medical imaging. The GPU has gained significant popularity as powerful tools for high performance computing (HPC) because of the *low cost , flexible and accessible* of the GPU[22]. Figure 10 illustrates architectural differences between GPUs and CPUs , the GPU has a number of threads where each thread can execution different program.

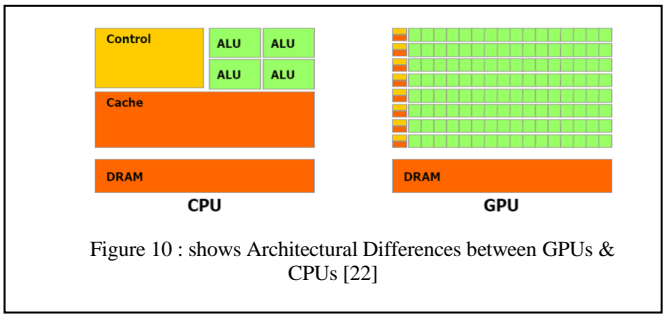

Figure 10 : shows Architectural Differences between GPUs & CPUs [22]

Finally, the architecture of GPUs can be classified as Single Instruction Multiple Data (SIMD) .

*3.2 Parallel Computing Toolbox in R2012b*
*First of all*, this version classified as the newest version of Matlab from Mathworks. It introduced in September 2012 . It has many new features such as : 1) supporting programming with CUDA NVIDIA for getting more advantages of GPUs; 2) supporting parallel for-loops *(parfor)* for running task-parallel algorithms on multiple processors ; 3) R2012b can be run 12 workers locally on a multicore desktop ; 4) for Matlab Distributed Computing Server (DCT) computer it supports cluster and grid; 5) interactive and batch execution of parallel applications ; and 6) distributed arrays and single program multiple data for large dataset handling and data-parallel algorithms[14]. Figure 10 shows how CPT R2012b run applications on a multicore desktop with 12 workers available in the toolbox, take advantage of GPUs, and scale up to a cluster with Matlab Distributed Computing Server [19],[23].

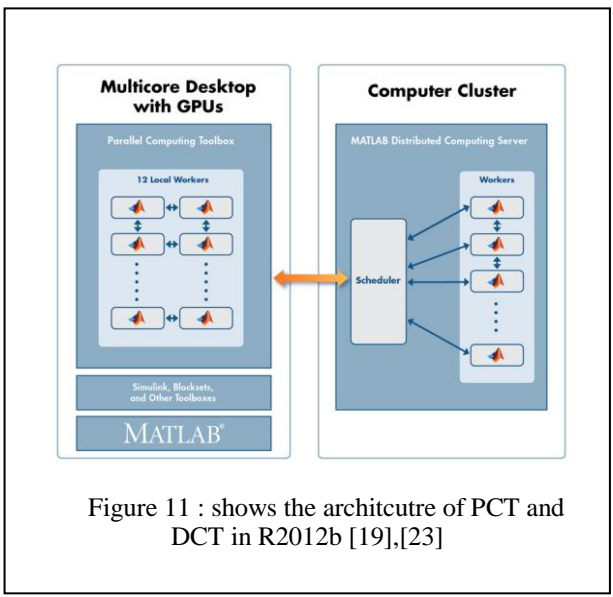

Figure 11 : shows the architcutre of PCT and DCT in R2012b [19],[23]

### 3.2(a) Benchmarking A\b on the GPU in R2012b

In this problem , we will show how can apply linear system on the GPU. Here we want to benchmark matrix left division (\),that means  x = A\b, using CPU and GPU . Hence , the code for this problem shows below:

```
Function results = paralleldemo_gpu_backslash(maxMemory)
```

Now the important thing is choosing the fitting matrix size . The appropriate matrix size based on system memory available in your system. The figure 12 show the compare the performance on the CPU and the GPU, both for single , double precision and speed up . The comparative performance of this problem is *Gigaflops* the number of floating point operations per second as our measure of performance because that allows us to compare the performance of the algorithm for different matrix sizes. For execution the benchmarks we will use this code.

```
function [gflopsCPU, gflopsGPU] = executeBenchmarks(clz, sizes)
    fprintf(['Starting benchmarks with %d different %s-precision ' ...
        'matrices of sizes\nranging from %d-by-%d to %d-by-%d.\n'], ...
        length(sizes), clz, sizes(1), sizes(1), ...
        sizes(end), sizes(end));
    gflopsGPU = zeros(size(sizes));
    gflopsCPU = zeros(size(sizes));
    for i = 1:length(sizes)
        n = sizes(i);
        [A, b] = getData(n, clz);
        gflopsCPU(i) = benchFcn(A, b);
        fprintf('Gigaflops on CPU: %f\n', gflopsCPU(i));
        A = gpuArray(A);
        b = gpuArray(b);
        gflopsGPU(i) = benchFcn(A, b);
        fprintf('Gigaflops on GPU: %f\n', gflopsGPU(i));
    end
end
```

As example , one iteration of creating matrix as below:
*Creating a matrix of size 16384-by-16384.*
*Gigaflops on CPU: 53.667451*
*Gigaflops on GPU: 260.272952*
For more details about this problem see [19].

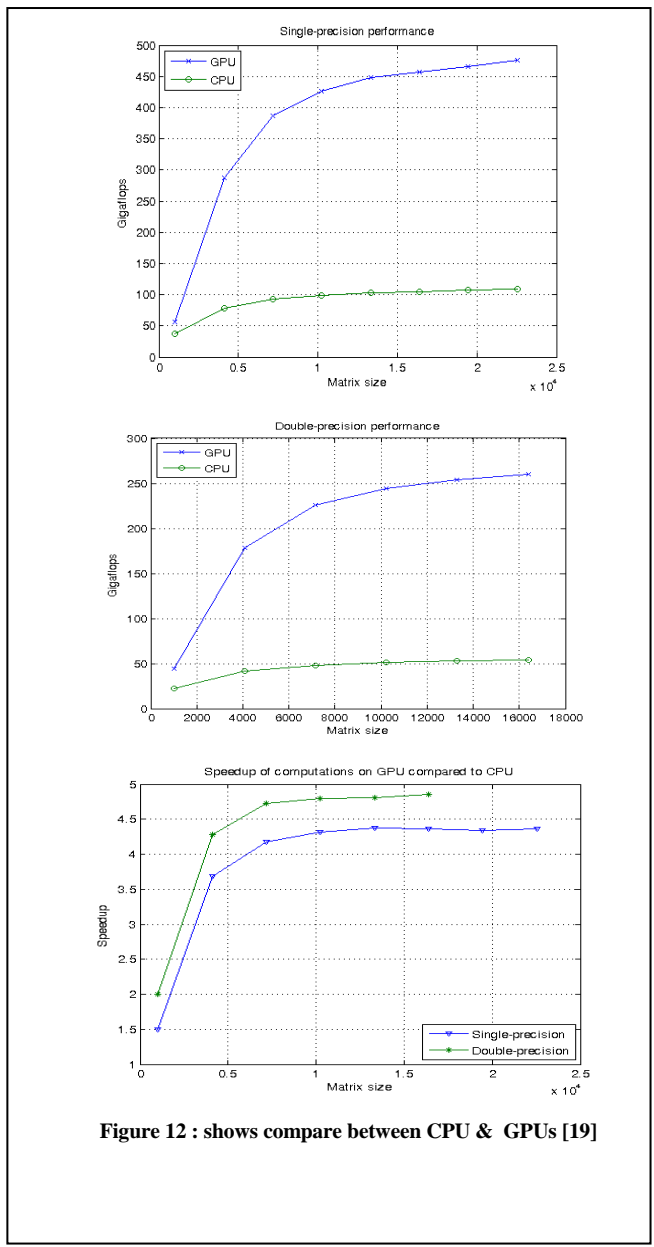

**Figure 12 : shows compare between CPU & GPUs [19]**

### 4. PARALLE MATALAB FOR NEAR FUTURE

In the near future, the author expects Matlab will be increasing the number of works in the next version of Parallel Computing Tools (PCT) to become more than 20 workers on a single machine in *R2014b*. Also Matlab will be more compatible with *CUDA* and it tries to improve the current support.

### 5. CONCLUSSIONS AND FUTURE WORK

Matlab is one of the most widely used mathematical computing environments in technical computing. It has many advantages like easy to learn and use . In this paper, we presented how parallel computation in Matlab can be immensely beneficial to researchers in many fields such as signal and image processing . As these provide avenues for fast and efficient computation, especially in modern hardware which has multicore processors. The authors presented the most attempts parallel Matlab in the past,present such as MatlabMPI, bcMPI, pMatlab, Star-P and PCT. At the end of this paper the author shows his opinion for parallel Matlab in the near future.

*For the future work*, we are planning to analysis and implement a new tool, which combines the advantages of previous versions to get the highest performance, ease and scalability of the tool. We suggested using open source programming such as GNU Octave [24] for the new tool .

#### ACKNOWLEDGMENT

The author would like to thank *Prof. Rosni Abdullah and Dr. Mohd. Adib Hj. Omar* from University Science Malaysia, in order to give me this opportunity to study attempts parallel Matlab.